\documentclass{evn2002}
\usepackage{graphicx}
\begin{document}
   \title{Understanding Scintillation of Intraday Variables}

   \author{T.\ Beckert,
           L.\ Fuhrmann,
           G.\ Cim\`o,
           T.~P.\ Krichbaum,
           A.\ Witzel \and
           J.~A.\ Zensus 
          }
   \authorrunning{Beckert et al.}
   \institute{Max-Planck-Institut f\"ur Radioastronomie, Auf dem H\"ugel 69,
              53121 Bonn, Germany
             }

   \abstract{
Intraday Variability of compact extragalactic radio sources can be interpreted
as quenched scintillation due to turbulent density fluctuations of the
nearby ionized interstellar medium. We demonstrate that the statistical
analysis of IDV time series contains both information about 
sub-structure of the source on the scale of several 10 $\mu$as
and about the turbulent state of the ISM. The source structure and
ISM properties cannot be disentangled using IDV observations alone.
A comparison with the known morphology of the `local bubble' and the
turbulent ISM from pulsar observations constrains possible source models. 
We further argue that earth orbit synthesis fails for  non-stationary
relativistic sources and no reliable 2D-Fourier reconstruction is
possible. 
   }

   \maketitle
%

\section{Introduction}
Intraday Variability (IDV) (see Wagner \& Witzel (\cite{Wagner}) for a review) 
is a common phenomenon of flat-spectrum radio cores in quasars and BL Lacs.
From light travel time arguments the observed brightness temperature
of IDV sources are in the range of $10^{16 \ldots 21}$~K and far in
excess of the inverse Compton limit. The suggested intrinsic explanation
require either extreme Doppler boosting or special source geometries.
Gravitational micro-lensing and
scintillation in the ISM of our galaxy are also discussed as possible 
propagation effects causing variability.

Two classes of IDV sources\footnote{Besides normal IDV two extremely
fast sources PKS 0405-385 (Kedziora-Chudczer et al.
\cite{Kedziora-Chudczer}) and
J1819+3845 (Dennett-Thorpe \& de Bruyn \cite{Dennett-Thorpe}) 
have been found with very short time-scales of less than $1$ hour.}
are distinguished according to their 
structure functions ($S\!F$). One shows a continuous increase of 
$S\!F$ with time-lag, while the other reaches a `plateau' at a well 
defined time-scale $t_\mathrm{IDV}$ (see Fig.~1).
 

In this paper we will discuss the scintillation hypothesis for the 
`plateau' class.
Scintillation is caused by turbulence in the ISM, which is a stochastic 
process and the resulting time series are best analysed in terms of 
first order structure functions. Following Simonetti, Cordes \& Heeschen
(\cite{Simonetti}) we define the discrete structure function for time series
$f(t_i)$ by
\begin{equation}
  S\!F(\tau_j) = N^{-1}_{ij}\sum_{i=1}^{n} w(i) w(i+j) 
  \left[f(t_i) - f(t_i + \tau_j)\right]^2 \,.
\end{equation} 
Here 
$N_{ij}$ is the normalisation; $w(i)$ are weighting functions, so that 
$w(i) w(i+j) >0$, if a
measurement at $t_i$ and another measurement at $t_i + \tau_j$ were obtained.
The weighting function also accounts for bining of unevenly sampled data.
The structure function is related to the autocorrelation $\rho(\tau)$ by
$S\!F(\tau) = 2[\rho(0) - \rho(\tau)]$.

   \begin{figure}
   \centering
   \includegraphics[width=\columnwidth]{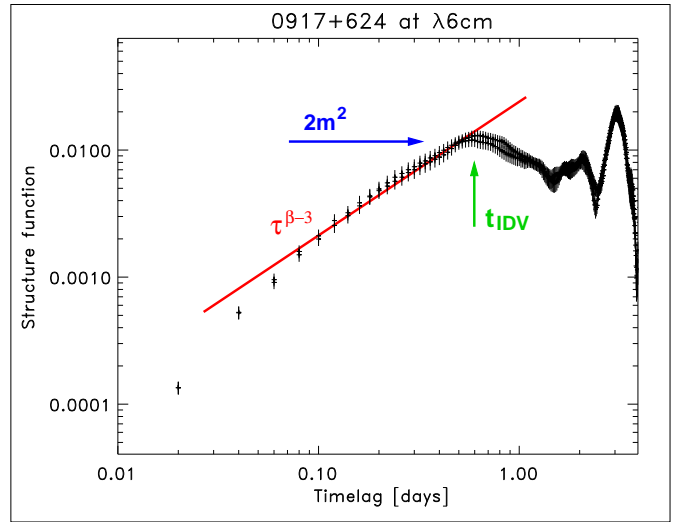}
      \caption{Structure function for IDV in total intensity for
               the quasar 0917+624. The $S\!F \propto \tau^{\alpha}$
               has a slope 
               of $\alpha \approx 1$, a 'plateau' with $S\!F = 2 m^2$ for long 
               time-lags and a characteristic timescale for decorrelation 
               $t_\mathrm{IDV}$ at the first maximum.
         \label{fig:0917_sf}
         }
   \end{figure}
%
\section{Quenched Scintillation}
Turbulence in the ionized ISM of our Galaxy produces density fluctuations
of free electrons and the corresponding refractive index for radio wave
propagation on a wide range of spatial scales (Armstrong, Rickett \& Spangler
\cite{Armstrong}). The power spectrum of density fluctuations in the ISM at
wave number $q$ along the line of sight (coordinate $z$) towards 
the quasar is described by a power law $\Phi(z,q) = C_N^2(z) q^{-\beta}$
between a lower and upper cut-off 
$q_\mathrm{low} < q < q_\mathrm{high}$. The lower cut-off in $q$ corresponds 
to the driving wavelength of the turbulent cascade. The inertial range 
described by the power law ends at $q_\mathrm{high}$, where energy 
is dissipated on small spatial scales.

One distinguishes refractive and diffractive
scattering at low frequencies and weak scattering above a characteristic
transition frequency $\nu_T$ (e.g. Narayan \cite{Narayan}). For quasars at high 
galactic latitude, the transition is expected between $1$ and $5$\,GHz
(e.g. Walker \cite{Walker}).
The angular scale for destructive interference of wavefronts due to optical path
length differences is the Fresnel scale
\begin{equation}
 \label{fresnel}
 \theta_F = \sqrt{\frac{\lambda}{2\pi z}} = 0.26 \mathrm{mas}\ 
 \left[\frac{\nu}{\mathrm{GHz}}\cdot \frac{z}{\mathrm{pc}}\right]^{-1/2}\;.
\end{equation}
Large fluctuations on scales smaller than $\theta_F$ leads to interference and 
diffractive scintillation, while fluctuations on scales larger than $\theta_F$
act as gradients in the refractive index and causes 
focusing and defocusing of light rays. 
     
Due to the 'large' size of incoherent extragalactic synchrotron sources 
\begin{equation} \label{thetaTb}
  \theta = 2.1 \mathrm{mas}\ \frac{(1+\tilde{z})}{\delta} 
  \left[\frac{\nu}{\mathrm{GHz}}\right]^{-1} 
  \left[\frac{F_\nu}{\mathrm{Jy}} \cdot 
  \frac{ 10^{11}\ \mathrm{K}}{T_b}\right]^{1/2}
\end{equation}
diffractive effects are unimportant, because the source size is always 
larger than the Fresnel scale for scattering in the ISM. Here $\tilde{z}$ is 
the cosmological redshift, $T_b$ the brightness temperature, $F_\nu$ the 
observed flux density of the source, $\nu$ the observing frequency,
and $\delta$  the Doppler boosting factor.  

For a scattering screen lying beyond the solar system and at
frequencies $\nu \ge 1$\,GHz the source size $\theta$ is always much larger
than the Fresnel scale and the amplitude of variations is quenched.
The amplitude of variations is reduced and the typical variability time-scale 
is stretched by the large size of the source.

The fluctuations in the ISM at scale $q$ change with a characteristic
turbulent velocity $\tilde{v}(q)$, which is assumed much smaller than the
relative velocity of earth and ISM. The scintillation pattern of intensity
variations appears to be frozen in the ISM and the spatial variations are
scanned by earth motion. The relative velocity of earth and ISM 
projected onto the sky has contributions from earth orbital motion,
$v_\mathrm{Earth}$, the velocity of the solar system in the local standard
of rest, $v_\mathrm{LSR}$, and the peculiar velocities of clouds and sheets
in the ISM, $v_\mathrm{cloud}$, like those in the wall of the local bubble
\begin{equation}
 v = v_\mathrm{Earth} + v_\mathrm{LSR} + v_\mathrm{cloud}\,.
\end{equation}     
Due to orbital motion, the velocity, $v$, traces an off-centred ellipse
in the plane of the sky. The velocity transfers variations on spatial
scales $x$ to temporal variations $\tau = x/v$ and leads to
an annual modulation of the variability time-scale 
(Rickett et al. \cite{Rickett1}).

From theoretical considerations (Coles et al. \cite{Coles}) the
autocorrelation $\rho(\tau)$ of light curves due to quenched scintillation can
be calculated as an integral along the line of sight through the
scattering medium 
of a Fourier transform $\vec{q}\rightarrow \vec{x}$ 
\begin{equation} 
  \label{autoc}
  \rho(\vec{x}) = 8 \pi r_e^2 \lambda^2 \int \mathrm{d}z\,
   \mathrm{FT}_{x\cdot q} \left\{ 
   \sin^2\left[ \vartheta_q^2 \right] \cdot \Phi(z,q) \cdot
   \left| V(\vec{q})\right|^2 \right\}
\end{equation}
of the geometric sinusoidal Fresnel term, the fluctuation power spectrum,
and the visibility of the source surface brightness $V(q)$.
The argument of the sinusoidal Fresnel term 
\begin{equation} 
 \vartheta_q = \frac{q z}{\sqrt{2}}\theta_F
\end{equation}
contains the Fresnel scale from (\ref{fresnel}).
The Fourier transform relates the 2-D vector $\vec{x}$
in the plane of sky, given by the vector of the velocity $\vec{v}$ to 
the 2-D wave vector $\vec{q}$ in the scattering medium.
In (\ref{autoc}) we have neglected a refractive cut-off in Fourier space for 
large $q$. For quenched scintillation the first relevant cut-off in $q$ is in the
visibility $V(\vec{q})$.

Anisotropic turbulence can be accounted for in $\Phi(z,\vec{q})$ but is not 
considered here. Non-circular and multi-component models for the source enter
(\ref{autoc}) through angular variations of the visibility in the 
$\vec{q}$-plane.  
\section{The Local ISM}
The ISM of the solar neighbourhood is structured in cloudlets, sheets, and walls.
The most prominent structure is the local bubble (e.g. Breitschwerdt, 
Freyberg \& Egger \cite{Breitschwerdt})
with a mean radius of $\sim 80$\,pc. Perpendicular to the galactic plane 
this bubble extends out to about 200\,pc, but its boundary is uncertain 
due to instabilities in the wall surrounding the bubble.   
The immediate vicinity of our solar system contains only small amounts of
dust and a low density ionized phase out to some tens of pc. Evidence for
absorption in the EUV of white dwarfs gives a mean density of
0.04 atoms/cm$^3$, an average bubble radius of $\sim 80$ pc,
and a five-fold increase in the gas density at the bubble
boundary (Warwick et al. \cite{Warwick93}).

The amplitude
of turbulence in the coronal gas of the local bubble is an order of magnitude
lower than elsewhere in the ISM. This is deduced from scintillation studies
of pulsars, which show reduced scintillation for the nearby pulsar 0950+08,
located near the edge of the local bubble, relative to other line of sights
through the ISM (Phillips \& Clegg \cite{Phillips}).

While quenched scintillation of extended sources favours contributions from
the nearby ISM (see (\ref{slab_m}) below), the scattering material
is probably associated with the wall of the local bubble. In interactions
with neighbouring bubbles, radio loops are produced and the bubble
interface is  Rayleigh-Taylor unstable.
This may have generated the large amplitude turbulence in the
ionized high density gas, necessary for scintillation.   
For scattering in localized regions like the wall of the local bubble 
the scattering measure 
\begin{equation}
S\!M = \int_{D-H/2}^{D+H/2} \mathrm{d}z\, C_N^2(z)
\end{equation}
can be separated in (\ref{autoc}) and the medium can be treated as a 
slab of thickness $H$ at distance $D$.

\section{The Source Structure}
The relation (\ref{autoc}) between autocorrelation of intensity
fluctuations and source visibility opens the path to an intensity
interferometer (Hanbury Brown \& Twiss \cite{Hanbury}). Unlike 
other interferometric techniques in the radio band, all phase information
is lost, because scintillation in quenched scattering does
not rely on constructive and destructive interference of disturbed
wavefronts.   
The appearance of the Fresnel term and the steep power spectrum of turbulence
in (\ref{autoc}) separates the `IDV interferometer' from the optical intensity 
interferometer and limits the angular scale to a small window in $\theta$ 
at which (\ref{autoc}) can be used.

The time series of intensity fluctuations samples
the autocorrelation $\rho(x)$ for different baseline lengths, $x$,
in the sky but only along a given direction due to earth motion. It is 
therefore impossible to solve the inverse problem of (\ref{autoc})
to get the source visibility $V(q)$ from single time series of a
few days or weeks.  

Even the comparison of source models and the resulting theoretical
$\rho(x)$  with observed IDV is hampered by our insufficient knowledge
of turbulence in the ISM. Any measurement of size and structure of the 
extragalactic source is affected by uncertainties in distance, scattering measure 
and relative velocity of the scattering medium. 

From (\ref{autoc}) we can derive the square of the modulation index
$m^2 = \lim_{\tau \rightarrow 0} \rho(\tau)$ of flux variations
with a Gaussian brightness distribution for the source of width
$\theta$:
\begin{equation}
  \rho(0) = m^2 =  \left(\frac{r_e}{D \theta^2}\right)^2
 \lambda^4 (D \theta)^{\beta-2}  S\!M \cdot \frac{\Gamma (3 - \beta/2)}{2} 
\end{equation}
But it turns out that the structure function
 \begin{equation}
  S\!F(\tau) = 2\rho(0) \left(\frac{v \tau}{D \theta}\right)^2
  \frac{3 - \beta/2}{4} 
\end{equation}
rises for small time-lags as 
$ S\!F \propto \tau^2$ independent of the spectral index $\beta$.
The observed $S\!F$
show always a much shallower rise (see Fig.1) with 
$S\!M \propto \tau^{\alpha}\,;\;
\alpha = 0.7 \ldots 1.2$ and are inconsistent with the Gaussian model. 
The variability time-scale
of IDV $t_{\mathrm{IDV}}$ derived from the first maximum of $S\!F$ gives
the source size $\theta \approx v\,t_{\mathrm{IDV}}/(1.2\,D)$
independent of the scattering measure.  

Another (extreme) source model is a disk of size $\theta$ with constant
surface brightness. In this case we find  
\begin{equation} \label{slab_m}
  \rho(0) = m^2 = 2 \left(\frac{r_e}{D \theta^2}\right)^2 
 \lambda^4 (D \theta)^{\beta-2} S\!M  \cdot F_1(\beta)\ .
\end{equation}
The special functions $F_1(\beta)$ \& $G_1(\beta)$
are of order unity\footnote{ 
The functions $F_1(\beta)$ \& $G_1(\beta)$ are characteristic for a 
model with constant surface brightness and depend only on the spectral 
index $\beta$: 
\begin{equation}
  F_1(\beta) = \frac{\Gamma (3 - \beta/2)\;\Gamma ((\beta -3)/2)}
   {\sqrt{\pi}\; \Gamma ((\beta - 2)/2)\; \Gamma (\beta/2)} \,; 
\end{equation}
\begin{equation}
  G_1(\beta) = \frac{2^{4-\beta}}{\pi (\beta-3)}
  \frac{\Gamma ((5 - \beta)/2))}{\Gamma ((\beta -1)/2)}  \,.
\end{equation}
}.
For a succession of slabs at different distances the slab closest to
the observer dominates the variations according to
$m^2 \propto S\!M D^{\beta -4}$ for power spectra  $\beta < 4$
and for sources larger than the Fresnel scale in the relevant slab.

The $S\!F$ for time-lags much smaller than the correlation
time $t_\mathrm{IDV}$ rises in this case as
\begin{equation}
  S\!F(\tau) = 2\rho(0) \left(\frac{v \tau}{D \theta}\right)^{\beta-3} 
  \frac{G_1(\beta)}{F_1(\beta)}
\end{equation}
The characteristic time-scale $t_{\mathrm{IDV}}$ of IDV derived
from $S\!F$ gives the source size 
\begin{equation} \label{size}
  \theta = \frac{v}{D} \frac{t_{\mathrm{IDV}}}{2}
\end{equation}
in this model.

The expected intrinsic brightness temperature $T_\mathrm{b}$ of extragalactic
flat-spectrum radio cores is expected to be between the 
equipartition temperature of
$4.7\ 10^{10}$\ K and the inverse Compton limit of about $6.7\ 10^{11}$\ K.
For a given $T_\mathrm{b}$ we can estimate the distance to the
scattering medium from (\ref{size}) to be
\begin{equation}
  D = 137 \mathrm{pc} \left[\frac{\nu}{5 \mathrm{GHz}}\right]
  \left[\frac{T_\mathrm{b}}{10^{11}\mathrm{K}}\right]
  \left[\frac{\delta}{10}\right]
  \left[\frac{F_\nu}{1 \mathrm{Jy}}\right]^{-1/2}\, .
\end{equation}
where we assumed a velocity of $v = 20$\ km/s and $t_{\mathrm{IDV}} = 1$~day.
$\delta$ is the usual Doppler boosting factor and $F_\nu$ the observed
flux density of the scintillating component.
This distance agrees well with the distance to
the wall of the local bubble.

   \begin{figure}
   \centering
   \includegraphics[width=\columnwidth]{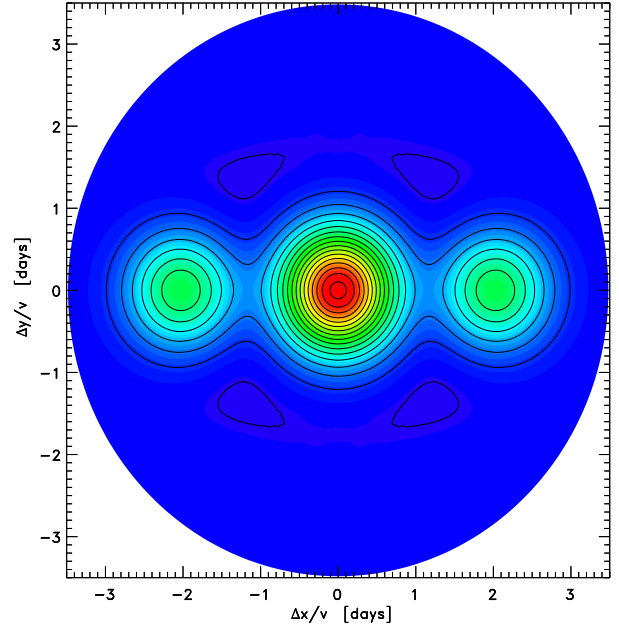}
      \caption{Autocorrelation of scintillation pattern for a two component
      source model. The components have Gaussian surface brightness width
      $\theta = 50 \mu$as. The separation is $7 \theta$.
         \label{fig:autoc}
         }
   \end{figure}
%

   \begin{figure}
   \centering
   \includegraphics[width=\columnwidth]{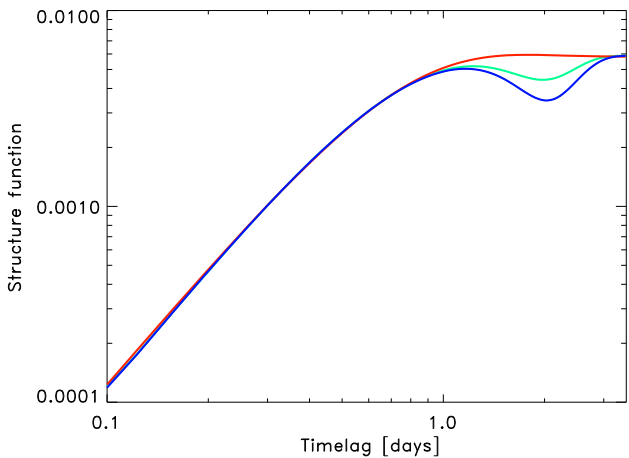}
      \caption{Structure function for a two component
      source model. The source model is like the one in Fig.\,\ref{fig:autoc}.
      The $S\!F$ are calculated for motion along (solid), perpendicular
      (dashed), and at $14^\circ$ to the jet axis.
         \label{fig:sf_two}
         }
   \end{figure}
%

So far we have assumed the source to be spherically symmetric despite
the fact that  most IDV sources have one-sided jets.  

An indication of deviation from spherical symmetry is seen in Fig.~1 where
the `plateau' in the structure function is not flat but has at
least one minimum
beyond $t_\mathrm{IDV}$. Besides anisotropy of turbulence in the ISM, 
a two-component source structure can lead to this effect as seen in Fig.~3.
The autocorrelation $\rho(\vec{x})$ of a two-component model
with two components of $\theta = 50\mu$as each and the same flux,
separated by $\theta_2 = 7 \theta$ is shown in Fig.~2.
$\rho(\vec{x})$ has two symmetric secondary maxima along the
line connecting the two components in the sky. In such a model the minima in 
$S\!F$ appear only when the velocity $\vec{v}$ is within $\pm 15^\circ$ of the
`jet-axis' connecting the two components as seen from Fig.~3.  
    
\section{Earth Orbit Synthesis? \label{synthesis}}
Macquart \& Jauncey (\cite{Macquart}) point out that the 
degeneracy of aligned baselines is broken by changes of earth
orbit direction. Within 6 months the projected earth motion
scans $180^\circ$ of the $\vec{v}$ ellipse in the sky.  
In that way the plane of $\rho(\vec{x})$ like the one in Fig.~2 can be 
sampled along several directions.
In principle the Fourier transform of (\ref{autoc}) can then be inverted. 
Besides knowledge of ISM properties as mentioned above, the source must
not suffer structural changes within at least 3 months. 

In the course of 3 months the distant quasar can undergo
structural changes.  Flat spectrum core-dominated radio
sources are certainly jet-sources and emit boosted synchrotron 
radiation.
Structural changes and flux variations of the source are limited by the
light travel time for the source radius\footnote{$d_A$ is the
angular diameter distance. For the cosmological parameters
we use $H_0 = 70$ km/s Mpc$^{-1}$ and $q_0 =0.5$.}
$d_A \theta$. For a source at a medium distance $\tilde{z}=1$ we get a
lower limit on the variability time of
\begin{equation}
  t_{\mathrm{var}} \le 122 \mathrm{days} 
  \left[\frac{\delta}{10}\right]^{-2}
  \left[\frac{\nu}{5 \mathrm{GHz}}\right]^{-1}
  \left[\frac{F_\nu}{1 \mathrm{Jy}}\cdot
  \frac{10^{11}\mathrm{K}}{T_\mathrm{b}} \right]^{1/2}
\end{equation}
which is about 4 months and therefore comparable to the required synthesis
time.

\section{Conclusions}
Quenched scintillation gives a consistent picture of Intraday Variability for
most IDV sources. The time-scale of about 1~day fits with scattering
in the ionized gas of the wall of the local bubble of a
moderately boosted synchrotron source with brightness temperatures below the
inverse Compton limit.

From the analytic expressions for modulation index $m$ and structure function
at small time-lags presented here, it is possible to determine 
two of the four unknowns: distance to the scattering medium, 
scattering measure, velocity of the medium, and size of the source.
In the slab model of the scattering medium together with a constant 
surface brightness of the source it is possible to determine the slope
$\beta \approx 4$ of turbulent power spectrum from the structure function. 

The structure of IDV sources beyond  the size of the most compact component
cannot be uniquely determined from existing 
experiments. Because most IDV sources are intrinsically variable on
time-scales of several months earth orbit synthesis cannot be
reliably realized.

Extreme scattering events (Fiedler et al. \cite{Fiedler}) show that
the ISM is structured
on scales of several AU down to 0.01 AU (Cim\`o et al. \cite{Cimo}) well into
the scintillation domain discussed here. Turbulence in the ISM 
is therefore not homogeneous on scales of $v$ (3 months) = 1~AU. Changes of the
long term variability pattern like the sudden end of
IDV in 0917+624 (Fuhrmann et al. \cite{Fuhrmann}) can be due to changes
intrinsic to the source or in the scattering measure of the ISM.

\end{document}